\documentclass[aps,prl,10pt,twocolumn,superscriptaddress,nofootinbib,showkeys,showpacs,altaffilletter]{revtex4-1}

\usepackage{graphicx}
\usepackage{dcolumn}
\usepackage{amssymb}
\usepackage{amsmath}
\usepackage{amsfonts}
\usepackage{amsbsy}
\usepackage{color}
\usepackage{rotating}
\usepackage[english]{babel}
\usepackage{soul}

\usepackage{hyperref}
\hypersetup{
    colorlinks=true,
    linkcolor=blue,
    filecolor=magenta,
    urlcolor=cyan,
    citecolor=cyan
}

\newcommand{\be}{\begin{equation}}
\newcommand{\ee}{\end{equation}}
\newcommand{\bea}{\begin{eqnarray}}
\newcommand{\eea}{\end{eqnarray}}

\begin{document}

\title{Geometrical observational bounds on a fractal horizon holographic dark energy}

\date{\today}

\author{Mariusz P. D\c{a}browski}
\email{mariusz.dabrowski@usz.edu.pl}
\affiliation{Institute of Physics, University of Szczecin, Wielkopolska 15, 70-451 Szczecin, Poland}
\affiliation{National Centre for Nuclear Research, Andrzeja So{\l}tana 7, 05-400 Otwock, Poland}
\affiliation{Copernicus Center for Interdisciplinary Studies, Szczepa\'nska 1/5, 31-011 Krak\'ow, Poland}
\author{Vincenzo Salzano}
\email{vincenzo.salzano@usz.edu.pl}
\affiliation{Institute of Physics, University of Szczecin, Wielkopolska 15, 70-451 Szczecin, Poland}

\begin{abstract}
A novel fractal structure for the cosmological horizon, inspired by COVID-19 geometry, which results in a modified area entropy, is applied to cosmology in order to serve dark energy. The constraints based on a complete set of observational data are derived. There is a strong Bayesian evidence in favor of such a dark energy in comparison to a standard $\Lambda$CDM model and that this energy cannot be reduced to a cosmological constant. Besides, there is a shift towards smaller values of baryon density parameter and towards larger values of the Hubble parameter, which reduces the Hubble tension.
\end{abstract}


\maketitle

\section{Introduction}

Black holes and cosmological horizons are very strongly explored phenomena in physics recently because they make the link between the classical and the quantum in the context of gravity. Through the Hawking temperature \cite{Hawking:1974sw} and Bekenstein area entropy \cite{Bekenstein:1974ax}, they allow thermodynamics to be related to the classical geometry of space. So, any modification of geometry will influence the entropy related to the horizons. A modification of fractal nature has been recently proposed by Barrow \cite{Barrow:2020tzx} who was inspired by the COVID-19 virus geometrical structure. The idea is to consider the core sphere of the horizon with the attached number $N$ of heavily packed smaller spheres, to each of which, step by step, a number $N$ of some smaller spheres are latched on and so on, forming a fractal. The obtained fractal, which contains a series of smaller and smaller spheres, is analogous to known fractal structures such as the Koch snowflake, the Sierpi\'nski gasket, and the Menger sponge \cite{Barrow:2020tzx} - one can find many of their graphical representations in the literature and over the internet. Simple calculations which apply geometrical series allow to add up all the surfaces of a hierarchical system of spheres which can be called the ``sphereflake''. The recurrence formula for the radius $r_{n+1}$ of the $(n+1)$-th sphere is $r_{n+1} = \lambda r_n$, where $r_n$ is the radius of the $n$-th sphere, and $\lambda <1$. Taking infinite number of steps $(n \to \infty)$, we can count up the effective surface area of all the spheres as a geometrical series
\be
A_{eff} = \sum_{n=0}^{\infty} N^n 4\pi (\lambda^n r)^2
= \frac{4\pi r^2}{1 - N\lambda^2} = 4\pi r_{eff}^2
\ee
with an effective radius being defined as
\be
r_{eff} = \frac{r}{\sqrt{1-N\lambda^2}} \equiv r^{1+\Delta/2},
\ee
where $0 \leq \Delta \leq 1$, provided that $N\lambda^2 <1$. The resulting effective surface area $A_{eff} \propto r_{eff}^2$ is larger than the core sphere surface area $A \propto r^2$ (so $r \propto A^{1/2}$) of radius $r$ and in the extreme case of $\Delta =1$, it acts as if it was a volume because the spheres cover a piece of a 3-volume, despite they are surfaces. In other words, geometrically, we sum up the areas, but since our main concern is the area entropy issue, then we deal with information aspects of the problem. In fact, when $\Delta = 1$ one has the most intricate surface of the horizon with the COVID-19-like fractal geometry. Modification of the horizon area immediately leads to a change of the effective Bekenstein entropy, making it larger than in a smooth case, according to the formula
\begin{equation}
S_{eff} \propto A_{eff} \propto r_{eff}^2 \propto r^{2 + \Delta} \propto  A^{1+ \frac{\Delta}{2}} ,
\end{equation}
where we can take as the smooth core sphere radius $r$ either the black hole Schwarzschild radius $r_s$ or the cosmological horizon length $L$. The exact formula in terms of the Planck area $A_{Pl}$ is given by \cite{Barrow:2020tzx}
\be
S_{eff} = k_B \left( \frac{A}{A_{Pl}} \right)^{1 + \frac{\Delta}{2}} ,
\ee
where $k_B$ is the Boltzmann constant.

In cosmology, there has been a vivid discussion of possible explanations of the dark energy phenomenon as contributed from cosmological horizons, leading to holographic dark energy  \cite{Wang:2016och}. It emerges that the fractal horizon can extra contribute to the matter. This has been first calculated in \cite{Saridakis:2020zol} and then tested by observational data from Type Ia Supernovae (SNeIa) Pantheon sample and passively evolving early-type galaxies acting as Cosmic Chronometers (CC) in \cite{Anagnostopoulos:2020ctz}.

In this work we apply the full set of cosmological data up-to-date to derive the most comprehensive bounds on the holographic and fractal parameters. The issue is that they strongly differ from the ones obtained in \cite{Saridakis:2020zol}, as we will show later.

\section{Theoretical background}

According to \cite{Wang:2016och} the holographic dark energy is given by $\rho_{H} \propto S_{eff} L^{-4}$ with the effective Bekenstein entropy $S_{eff} \propto A_{eff} \propto L^{2+\Delta}$, where $L$ is the horizon length. Thus, following \cite{Saridakis:2020zol} we can express Barrow holographic dark energy (BH) as:
\begin{equation}\label{eq:BH_dens}
\rho_{BH} = \frac{3\,C^2}{8\pi G} L^{2\left(\frac{\Delta}{2}-1\right)}\, ,
\end{equation}
where $C$ is the holographic parameter with dimensions of $[\mathsf{T}]^{-1}[\mathsf{L}]^{1-\Delta/2}$ and $G$ the Newton gravitational constant. As suggested in \cite{Li:2004rb}, we identify the length $L$ with the future event horizon:
\begin{equation} \label{eq:event_hor}
L \equiv a\, \int^{\infty}_{t} \frac{dt'}{a} = a \int^{\infty}_{a} \frac{da'}{H(a')a'^{2}} \, ,
\end{equation}
where $a$ is the scale factor. The cosmological equation is simply
\begin{equation} \label{eq:fried_eq}
H^{2} = \frac{8 \pi G}{3} \left(\rho_{m} +\rho_{r} +\rho_{BH}\right)\, ,
\end{equation}
where the suffices $m$ and $r$ refer respectively to matter and radiation. Note that the standard continuity equation for matter and radiation is still valid, i.e.
\begin{equation}
\dot{\rho}_{m,r} + 3 H \left(\rho_{m,r} + \frac{p_{m,r}}{c^{2}} \right) = 0\, ,
\end{equation}
where the pressure $p_{i} = w_{i} \rho_{i}$, with the equation of state parameter $w_{i}$ being $0$ for standard pressureless matter and $1/3$ for radiation. We rewrite Eq.~(\ref{eq:fried_eq}) as
\begin{equation}\label{eq:fried_eq_n1}
1 = \Omega_m(a) + \Omega_r(a) + \Omega_{H}(a)\, ,
\end{equation}
introducing the dimensionless density parameters $\Omega_{i}(a)$\footnote{We will follow the standard convention for which $\Omega_i(a)$, as a function, is the dimensionless density parameter at any scale factor $a$, while $\Omega_{i}$, with no argument specified, will be the dimensionless density parameter at the present time, i.e. $\Omega_i(a=1)= \Omega_{i}$.}, defined as
\begin{equation}\label{eq:dimensionless_dens}
\Omega_{m,r}(a) = \frac{H^{2}_0}{H^{2}(a)} \Omega_{m,r} a^{-3(1+w_{m,r})} \,, \\
\end{equation}
\begin{equation}\label{eq:dimensionless_dens_BH}
\Omega_{BH}(a) = \frac{C^2 }{H^{2}(a)}  L^{2 \left(\frac{\Delta}{2}-1\right)}\,.
\end{equation}
Combining Eqs.~(\ref{eq:fried_eq_n1})~-~(\ref{eq:dimensionless_dens}) and (\ref{eq:dimensionless_dens_BH}) one can express the Hubble parameter as
\begin{equation} \label{eq:Hubble_alt}
H(a) = H_{0} \sqrt{\frac{\Omega_{m} a^{-3} + \Omega_{r} a^{-4}}{1-\Omega_{BH}(a)}}\, .
\end{equation}
In order to find the evolution of the holographic dark energy density, we follow \cite{Li:2004rb} and \cite{Saridakis:2020zol} procedure: \textit{(1.)} we insert Eq.~(\ref{eq:Hubble_alt}) into Eq.~(\ref{eq:event_hor}); \textit{(2.)} we obtain the future event horizon from inversion of  Eq.~(\ref{eq:dimensionless_dens_BH}); \textit{(3.)} we compare results from step (1) and step (2) and differentiate both of them w.r.t. to $a$. We end up with the following differential equation for the BH dark energy (where prime is derivative with respect to $a$):
\begin{eqnarray}
a \Omega'_{H}(a)&=& \left(1+\frac{\Delta}{2}\right)\mathcal{F}_{r}(a) \\
&+& \Omega_{H}(a) \left( 1 - \Omega_{H}(a) \right) \left[ \left(1+2 \frac{\Delta}{2}\right) \mathcal{F}_{m}(a)\right. \nonumber \\
&+& \left. \left(1-\Omega_{H}(a)\right)^{\frac{\Delta/2}{2\left(\frac{\Delta}{2}-1\right)}} \Omega_{H}(a)^{\frac{1}{2\left(1-\frac{\Delta}{2} \right)}}  \mathcal{Q}(a)\right]\, , \nonumber
\end{eqnarray}
with:
\begin{eqnarray}
\mathcal{F}_{r}(a) &=& -\frac{2 \Omega_r a^{-4}}{\Omega_m a^{-3} + \Omega_r a^{-4}}\, , \\
\mathcal{F}_{m}(a) &=& \frac{\Omega_m a^{-3}}{\Omega_m a^{-3} + \Omega_r a^{-4}} \nonumber\, , \\
\mathcal{Q}(a) &=& 2 \left(1-\frac{\Delta}{2}\right) \left(H_0 \sqrt{\Omega_m a^{-3} + \Omega_r a^{-4}} \right)^{\frac{\Delta/2}{1-\frac{\Delta}{2}}} C^{\frac{1}{\frac{\Delta}{2}-1}}\, . \nonumber
\end{eqnarray}
It is easy to verify that in the limit $\Omega_r \rightarrow 0 $ one retrieves Eq.~(14) from \cite{Saridakis:2020zol}.

Although the main focus of this work will be based on the above discussion, we need to emphasize that the identification of the boundary with the event horizon, Eq.~(\ref{eq:event_hor}), is not the only possible choice. Following \cite{Hsu:2004ri}, in \cite{Li:2004rb} it was assumed that only the event horizon could give rise to a proper accelerated expansion, i.e. to a holographic dark energy, while particle and Hubble horizons were unsuitable. But in \cite{Pavon:2005yx} it was shown that indeed the Hubble horizon can be assumed as boundary and can induce a proper holographic dark energy, but only if an interaction between dark energy and dark matter was allowed. We will show here that the BH model can instead have the Hubble horizon as a boundary and result in an effective dark energy behaviour without the need of any interaction between the dark sectors.

If we assume the Hubble horizon as the boundary, namely
\begin{equation}
L \equiv \frac{c}{H(a)}\, ,
\end{equation}
the BH dark energy density would look like (absorbing the speed of light $c$ into the constant $C$):
\begin{equation}
\rho_{BH} = \frac{3\,C^2}{8\pi G} H^{2\left(1-\frac{\Delta}{2}\right)}\, .
\end{equation}
Still relying on the separate conservation of both matter and radiation, Eq.~(\ref{eq:fried_eq_n1}) would remain the same, but with the dimensionless BH density parameter defined as
\begin{equation}
\Omega_{BH}(a) = \frac{C^2}{H^{2}}  H^{2\left(1-\frac{\Delta}{2}\right)} = C^2 H^{-\Delta}\,.
\end{equation}
We will now analyze what happens in the two extremal cases, i.e. $\Delta=0$ and $\Delta=1$. For $\Delta=0$, from Eq.~(\ref{eq:fried_eq_n1}) we find
\begin{equation}
\Omega_{BH}(a=1) = C^2 = 1 - \Omega_m - \Omega_r\, ,
\end{equation}
so that Eq.~(\ref{eq:fried_eq}) becomes:
\begin{equation}
H^2 = H^{2}_{0} \left( \Omega_m a^{-3} + \Omega_r a^{-4} \right) + (1-\Omega_m-\Omega_r) H^2\; .
\end{equation}
If we consider negligible the contribution of radiation we arrive to $H^2 \propto a^{-3}$, $\rho_{BH} \propto H^2 \propto a^{-3}$, $\Omega_m \propto const.$ and $\Omega_{BH} \propto const.$, which clearly reproduce the flaws highlighted in \cite{Li:2004rb} which can be solved only if we add interaction in the dark sector as suggested in \cite{Pavon:2005yx}.

\begin{figure}[ht!]
\centering
\includegraphics[width=0.45\textwidth]{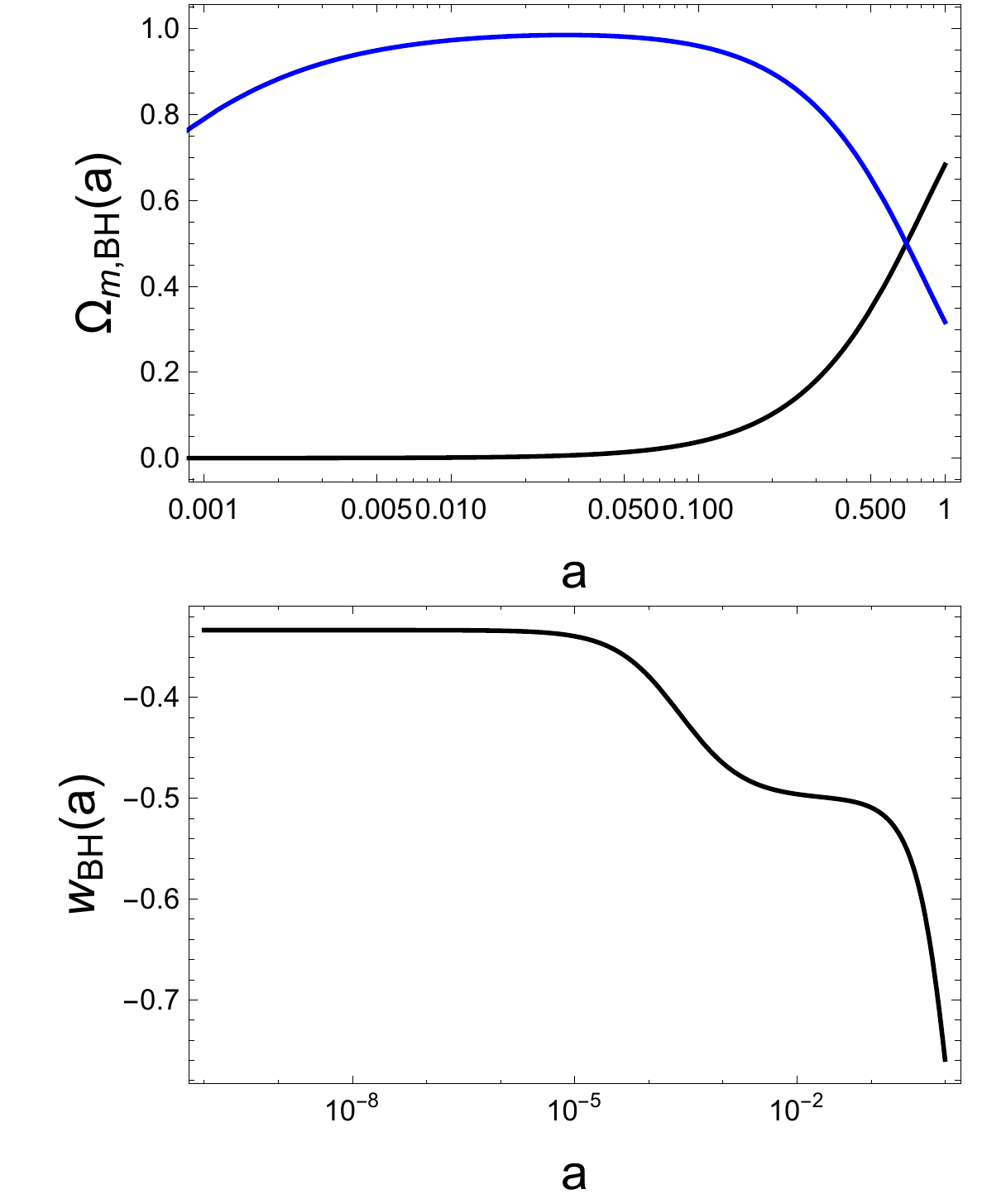}
\caption{Evolution with time of \textit{(top panel)} dimensionless density parameters for Barrow holographic dark energy, with the Hubble radius as boundary, $\Delta=1$, $\Omega_m = 0.317$, $\Omega_r = 8.4\cdot 10^{-5}$, $H_0=70.5$ km s$^{-1}$ Mpc$^{-1}$ and $C= 2$, and \textit{(bottom panel)} BH dark energy effective equation of state parameter. Colors: matter (blue), dark energy (black).}\label{plot:new}
\end{figure}

But $\Delta=0$ is a trivial case, where BH model is reduced to the standard one. The situation is totally different when $\Delta=1$; indeed, in this case from Eq.~(\ref{eq:fried_eq_n1}) we have
\begin{equation}
C^2 = \left( 1 - \Omega_m - \Omega_r \right) H_{0}\, ,
\end{equation}
and Eq.~(\ref{eq:fried_eq}) becomes a second order equation in the Hubble parameter:
\begin{equation}
H^2 - \left[(1-\Omega_m-\Omega_r)H_0 \right]H - H^{2}_{0} \left( \Omega_m a^{-3} + \Omega_r a^{-4} \right) = 0\; ,
\end{equation}
whose solution is:
\begin{eqnarray}
H(a) &=& \frac{H_0}{2} \left[(1-\Omega_m-\Omega_r)+  \right. \\
&& \left. \sqrt{(1-\Omega_m-\Omega_r)^2+4\left(\Omega_m a^{-3}+\Omega_r a^{-4}\right)}\right]\;. \nonumber
\end{eqnarray}
Now we cannot easily disentangle the behaviour of the effective holographic dark energy from the other components, but we can have an insight from studying the behaviour of the dimensionless density,
\begin{equation}
\Omega_{BH}(a) = \frac{(1-\Omega_m-\Omega_r)H_0}{H(a)}\, ,
\end{equation}
and of the effective equation of state parameter, $w_{BH}=p_{BH}/\rho_{BH}$, for the BH dark energy which can be derived from the standard continuity equation
\begin{equation}
\dot{\rho}_{BH} + 3 H \rho_{BH}(1+w_{BH}) = 0\, ,
\end{equation}
and in this case it results to be
\begin{equation}
w_{BH} = -1 -\frac{\dot{H}}{3H^2} = -1 -\frac{a H H'}{3H^2}.
\end{equation}
In Fig.~\ref{plot:new}, where we have $\Omega_m(a)$ in blue and $\Omega_{BH}(a)$ in black in the top panel, and $w_{BH}(a)$ in the bottom one, we clearly show that the Hubble radius, in the BH dark energy context, can safely play the role of boundary, implying a dark energy behaviour (note the limit $w_{BH}\xrightarrow[a \to 0]{}-1/3$ independent of the set parameters) without any need to introduce interaction in the dark sector.

\section{Statistical Analysis}

To analyze in full detail the compatibility of BH with cosmological data, we use the most updated set of data available today related to the geometrical global evolution of our Universe at large scales. We consider: Type Ia Supernovae (SNeIa) from the Pantheon sample; Cosmic Chronometers (CC); the gravitational lensing data from COSMOGRAIL's Wellspring project (H0LiCOW); the ``Mayflower'' sample of Gamma Ray Bursts (GRBs); Baryon Acoustic Oscillations (BAO) from several surveys; and the latest \textit{Planck} $2018$ release for Cosmic Microwave Background radiation (CMB).

We consider two different cases: the set which we call ``full data'', where we join both early- (CMB and BAO data from SDSS) and late-time observations (SNeIa, CC, H0LiCOW, GRBs and BAO data from WiggleZ); and the ``late-time'' data set, which includes only late-time data. We have decided to consider these two cases separately, because while on the one hand early-time data have much more stringent constraining power in cosmological model inference than late-time ones, on the other hand, they seem to be biased to statistically support a standard $\Lambda$CDM model, i.e. a cosmological constant as dark energy. By separating data in such a way, we could aim to have some more insight into a possible presence of a time varying dark energy candidate.

To perform our statistical analysis, we define the total $\chi^2$ as the sum of all the contributions considered, $\chi^{2}= \chi_{SN}^{2}+\chi_{G}^{2}+\chi_{H}^{2}+\chi_{H_{COW}}^{2}+ \chi_{BAO}^{2} + \chi_{CMB}^{2}$. To minimize the $\chi^2$ we use our own code implementation of a Monte Carlo Markov Chain (MCMC) \cite{Berg,MacKay,Neal} and we test its convergence using the method of \cite{Dunkley:2004sv}. Finally, we assess BH reliability using Bayesian Evidence, $\mathcal{E}$. Our reference model is the standard $\Lambda$CDM model, analyzed with the same set of data. Then, we calculate the Bayesian Evidence using the algorithm from \citep{Mukherjee:2005wg}. To reduce its prior dependence \citep{Nesseris:2012cq} and avoid any misleading estimation, we have used the same uninformative flat priors on the parameters for each model while running our MCMC codes, on a sufficiently wide range, so that a further increasing has negligible impact on $\mathcal{E}$. Such priors are mainly physically motivated: $0<\Omega_b < \Omega_m < 1$, $0<h<1$, $0\leq\Delta\leq1$ \cite{Barrow:2020tzx}, and $C>0$ (given Eq.~(\ref{eq:BH_dens}), we cannot discriminate among positive and negative values). After the Bayesian Evidence, we define the Bayes Factor as the ratio of evidence between two models, $M_{i}$ and $M_{j}$, $\mathcal{B}^{i}_{j} = \mathcal{E}_{i}/\mathcal{E}_{j}$: if $\mathcal{B}^{i}_{j} > 1$, model $M_i$ is preferred over $M_j$, given the data. As stated above, here the $\Lambda$CDM model will play the role of the reference models $M_j$. Finally, in order to state how much better is model $M_i$ with respect to model $M_j$, we have followed the Jeffreys' Scale \cite{Jeffreys:1939xee}.

\subsection{Type Ia Supernovae}

The Pantheon compilation \cite{Scolnic:2017caz} is made of $1048$ objects spanning the redshift range $0.01<z<2.26$. The corresponding $\chi^2_{SN}$ is defined as
\begin{equation}
\chi^2_{SN} = \Delta \boldsymbol{\mathcal{\mu}}^{SN} \; \cdot \; \mathbf{C}^{-1}_{SN} \; \cdot \; \Delta  \boldsymbol{\mathcal{\mu}}^{SN} \;,
\end{equation}
where $\Delta\boldsymbol{\mathcal{\mu}} = \mathcal{\mu}_{\rm theo} - \mathcal{\mu}_{\rm obs}$ is the difference between the theoretical and the observed value of the distance modulus for each SNeIa and $\mathbf{C}_{SN}$ represents the total covariance matrix. Note that we do not use the binned version as in \cite{Anagnostopoulos:2020ctz}, but the full one. The distance modulus is defined as
\begin{equation}
\mu(z,\boldsymbol{p}) = 5 \log_{10} [ d_{L}(z, \boldsymbol{p}) ] +\mu_0 \; ,
\end{equation}
where
\begin{equation}
d_L(z,\boldsymbol{p})=(1+z)\int_{0}^{z}\frac{dz'}{E(z',\boldsymbol{p})} \,
\end{equation}
is the dimensionless luminosity distance and $\boldsymbol{\theta}$ is the vector of cosmological parameters. Because of the degeneracy between the Hubble constant $H_0$ and the SNeIa absolute magnitude (both included in the nuisance parameter $\mu_0$), we marginalize the $\chi^{2}_{SN}$ over $\mu_0$ following \cite{conley}, obtaining
\begin{equation}\label{eq:chis}
\chi^2_{SN}=a+\log \left(\frac{d}{2\pi}\right)-\frac{b^2}{d},
\end{equation}
where $a\equiv\left(\Delta \boldsymbol{\mathcal{\mu}}_{SN}\right)^T \; \cdot \; \mathbf{C}^{-1}_{SN} \; \cdot \; \Delta  \boldsymbol{\mathcal{\mu}}_{SN}$, $b\equiv\left(\Delta \boldsymbol{\mathcal{\mu}}^{SN}\right)^T \; \cdot \; \mathbf{C}^{-1}_{SN} \; \cdot \; \boldsymbol{1}$, $d\equiv\boldsymbol{1}\; \cdot \; \mathbf{C}^{-1}_{SN} \; \cdot \;\boldsymbol{1}$ and $\boldsymbol{1}$ is the identity matrix.

\subsection{Cosmic Chronometers}

The definition of CC is used for Early-Type galaxies which exhibit a passive evolution and a characteristic feature in their spectra \cite{Jimenez:2001gg}, for which can be used as clocks and provide measurements of the Hubble parameter $H(z)$ \cite{Moresco:2010wh}. The sample we are going to use in this work is from \cite{moresco} and covers the redshift range $0<z<1.97$. The $\chi^2_{H}$ is defined as
\begin{equation}\label{eq:hubble_data}
\chi^2_{H}= \sum_{i=1}^{24} \frac{\left( H(z_{i},\boldsymbol{p})-H_{obs}(z_{i}) \right)^{2}}{\sigma^2_{H}(z_{i})} \; ,
\end{equation}
where $\sigma_{H}(z_{i})$ are the observational errors on the measured values $H_{obs}(z_{i})$.

\subsection{H0LiCOW}

On of the main novelty in the cosmological data set used here to constrain a cosmological model is the data from the H0LiCOW collaboration \cite{Suyu:2016qxx}. The main goal was to use the sensitivity of strong gravitational lensing events to constrain $H_0$ and, possibly, other parameters characterizing the cosmological background. H0LiCOW has used $6$ selected lensed quasars \cite{Wong:2019kwg} for which it was possible to retrieve multiple (lensing) images. Having multiple images makes it possible to take advantage of lensing time delay as a cosmological probes: in fact, it is well-known that during a lensing event, the light travel time from the sources (in this case, the quasars) to the observer depends both on the path length and on the gravitational potential of the foreground mass(ess) (the lenses). When multiple images are produced, they can exhibit a time delay at collection given by
\begin{equation}\label{eq:timedelGO}
    t(\boldsymbol{\theta},\boldsymbol{\beta})=\frac{1+z_L}{c}\frac{D_L D_S}{D_{LS}}\left[\frac{1}{2}(\boldsymbol{\theta}-\boldsymbol{\beta})^2-
    \hat{\Psi}(\boldsymbol{\theta})\right].
\end{equation}
In a typical gravitational lensing configuration \cite{gralen.boo}, $z_L$ is the lens redshift, $\boldsymbol{\theta}$ is the angular position of the image, $\boldsymbol{\beta}$ is the angular position of the source and $\hat{\Psi}$ is the effective lens potential. The distances $D_S$, $D_L$ and $D_{LS}$ are, respectively, the angular diameter distances from the source to the observer, from the lens to the observer, and between source and lens. The angular diameter distance is given by
\begin{equation} \label{eq:ang_dist}
D_{A}(z,\boldsymbol{p})=\frac{1}{1+z}\int_{0}^{z} \frac{c\, \mathrm{d}z'}{H(z',\boldsymbol{p})} \;,
\end{equation}
so that we have: $D_{S} = D_{A}(z_S)$, $D_{L} = D_{A}(z_L)$, and $D_{LS} = 1/(1+z_S) \left[(1+z_S)D_S - (1+z_L)D_L\right]$ \cite{Hogg:1999ad}. The combination of distances which appears in Eq.~(\ref{eq:timedelGO}),
\begin{equation}
D_{\Delta t} \equiv (1+z_L)\frac{D_L D_S}{D_{LS}}\, ,
\end{equation}
is generally called time-delay distance and is constrained by H0LiCOW. The data ($D^{obs}_{\Delta t,i}$) and the corresponding errors $(\sigma_{D_{\Delta t,i}})$ on this quantity for each of the $6$ considered quasars are provided in \cite{Wong:2019kwg}. Eventually, the $\chi^2$ for H0LiCOW data is
\begin{equation}\label{eq:cow_data}
\chi^2_{HCOW}= \sum_{i=1}^{6} \frac{\left( D_{\Delta t,i}(\boldsymbol{p})-D^{obs}_{\Delta t,i}\right)^{2}}{\sigma^2_{D_{\Delta t,i}}} \; ,
\end{equation}

\subsection{Gamma Ray Bursts}

Although the possibility to ``standardize'' GRBs is still on debate, we focus on the ``Mayflower'' sample, made of 79 GRBs in the redshift interval $1.44<z<8.1$ \cite{Liu:2014vda}, because it has been calibrated with a robust cosmological model independent procedure. The observational probe related to GRBs observable is the distance modulus, so the same procedure used for SNeIa is also applied here. The $\chi_{G}^2$ is thus given by $\chi^2_{GRB}=a+\log d/(2\pi)-b^2/d$ as well, with $a\equiv \left(\Delta\boldsymbol{\mathcal{\mu}}^{G}\right)^T \, \cdot \, \mathbf{C}^{-1}_{G} \, \cdot \, \Delta  \boldsymbol{\mathcal{\mu}}^{G}$, $b\equiv\left(\Delta \boldsymbol{\mathcal{\mu}}^{G}\right)^T \, \cdot \, \mathbf{C}^{-1}_{G} \, \cdot \, \boldsymbol{1}$ and $d\equiv\boldsymbol{1}\, \cdot \, \mathbf{C}^{-1}_{G} \, \cdot \, \boldsymbol{1}$.

\subsection{Baryon Acoustic Oscillations}

For BAO we consider multiple data sets from different surveys.  In general, the $\chi^2$
\begin{equation}
\chi^2_{BAO} = \Delta \boldsymbol{\mathcal{F}}^{BAO} \, \cdot \ \mathbf{C}^{-1}_{BAO} \, \cdot \, \Delta  \boldsymbol{\mathcal{F}}^{BAO} \ ,
\end{equation}
will have observables $\mathcal{F}^{BAO}$ which will change from survey to survey.

When we employ the data from the WiggleZ Dark Energy Survey (at redshifts $0.44$, $0.6$ and $0.73$) \cite{Blake:2012pj}, the relevant physical quantities are the acoustic parameter
\begin{equation}\label{eq:AWiggle}
A(z,\boldsymbol{p}) = 100  \sqrt{\Omega_{m} \, h^2} \frac{D_{V}(z,\boldsymbol{p})}{c \, z} \, ,
\end{equation}
where $h=H_0/100$, and the Alcock-Paczynski distortion parameter
\begin{equation}\label{eq:FWiggle}
F(z,\boldsymbol{p}) = (1+z)  \frac{D_{A}(z,\boldsymbol{p})\, H(z,\boldsymbol{p})}{c} \, ,
\end{equation}
where $D_{A}$ is the angular diameter distance defined in Eq.~(\ref{eq:ang_dist}) and
\begin{equation}
D_{V}(z,\boldsymbol{\theta})=\left[ (1+z)^2 D^{2}_{A}(z,\boldsymbol{\theta}) \frac{c z}{H(z,\boldsymbol{\theta})}\right]^{1/3}
\end{equation}
is the geometric mean of the radial $(\propto H^{-1})$ and tangential $(D_A)$ BAO modes. Note that this data set is independent of early-time evolution, thus it is included in the late-time data analysis.

We also consider data from multiple analysis of SDSS-III Baryon Oscillation Spectroscopic Survey (BOSS) observations. Each of the following data is used for the full data analysis but not for the late-time one.

In the DR$12$ analysis described in \cite{Alam:2016hwk}, the following quantities are given:
\begin{equation}
D_{M}(z,\boldsymbol{p}) \frac{r^{fid}_{s}(z_{d},)}{r_{s}(z_{d},\boldsymbol{p})}, \qquad H(z) \frac{r_{s}(z_{d},\boldsymbol{p})}{r^{fid}_{s}(z_{d})} \,,
\end{equation}
where the comoving distance $D_M$ is
\begin{equation}\label{eq:comovdist}
D_{M}(z,\boldsymbol{p})=\int_{0}^{z} \frac{c\, \mathrm{d}z'}{H(z',\boldsymbol{p})} \; ;
\end{equation}
the sound horizon evaluated at the dragging redshift is $r_{s}(z_{d})$; while $r^{fid}_{s}(z_{d})$ is the sound horizon calculated for a given fiducial cosmological model (in this case, it is $147.78$ Mpc). The dragging redshift is estimated using the analytical approximation provided in  \cite{Eisenstein:1997ik} as
\begin{equation}\label{eq:zdrag}
z_{d} = \frac{1291 (\Omega_{m} \, h^2)^{0.251}}{1+0.659(\Omega_{m} \, h^2)^{0.828}} \left[ 1+ b_{1} (\Omega_{b} \, h^2)^{b2}\right]\; ,
\end{equation}
where the factors $b_1$ and $b_2$ are given by
\begin{eqnarray}\label{eq:zdrag_b}
b_{1} &=& 0.313 (\Omega_{m} \, h^2)^{-0.419} \left[ 1+0.607 (\Omega_{m} \, h^2)^{0.6748}\right] \,,
	\nonumber \\
b_{2} &=& 0.238 (\Omega_{m} \, h^2)^{0.223}\,,
\end{eqnarray}
respectively. Finally, the sound horizon is defined as:
\begin{equation}\label{eq:soundhor}
r_{s}(z,\boldsymbol{p}) = \int^{\infty}_{z} \frac{c_{s}(z')}{H(z',\boldsymbol{p})} \mathrm{d}z'\, ,
\end{equation}
where the sound speed is given by
\begin{equation}\label{eq:soundspeed}
c_{s}(z) = \frac{c}{\sqrt{3(1+\overline{R}_{b}\, (1+z)^{-1})}} \; ,
\end{equation}
with the baryon-to-photon density ratio parameters defined as $\overline{R}_{b}= 31500 \Omega_{b} \, h^{2} \left( T_{CMB}/ 2.7 \right)^{-4}$, with $T_{CMB} = 2.726$ K.

From the DR$12$ we also include measurements derived from the void-galaxy cross-correlation \cite{Nadathur:2019mct}:
\begin{eqnarray}
\frac{D_{A}(z=0.57)}{r_{s}(z_{d})} &=& 9.383 \pm 0.077\,, \\
H(z=0.57) r_{s}(z_{d})  &=& (14.05 \pm 0.14) 10^{3} \, \rm{km\,s^{-1}} \, .
\end{eqnarray}

From the  extended Baryon Oscillation Spectroscopic Survey (eBOSS) we use the point $D_V(z=1.52)=3843\pm147\,r_s(zd)/r_s^{fid}(z_d)$ Mpc \cite{Ata:2017dya}.

Finally, we have also taken into account data from eBOSS DR14 obtained from the combination of the Quasar-Lyman $\alpha$ autocorrelation function \cite{lyman} with the cross-correlation measurement \cite{Blomqvist:2019rah}, namely
\begin{eqnarray}
\frac{D_{A}(z=2.34)}{r_{s}(z_{d})} &=& 36.98^{+1.26}_{-1.18}\,, \\
\frac{c}{H(z=2.34) r_{s}(z_{d})}  &=& 9.00^{+0.22}_{-0.22}\, .
\end{eqnarray}

\subsection{Cosmic Microwave Background}

As CMB data we use the shift parameters defined in \cite{Wang:2007mza} and derived from the latest \textit{Planck} $2018$ data release \cite{cmb&sn}. Thus, the $\chi^2_{CMB}$ is defined as
\begin{equation}
\chi^2_{CMB} = \Delta \boldsymbol{\mathcal{F}}^{CMB} \; \cdot \; \mathbf{C}^{-1}_{CMB} \; \cdot \; \Delta  \boldsymbol{\mathcal{F}}^{CMB} \; ,
\end{equation}
where the vector $\mathcal{F}^{CMB}$ is made of the quantities:
\begin{eqnarray}
R(\boldsymbol{p}) &\equiv& \sqrt{\Omega_m H^2_{0}} \frac{r(z_{\ast},\boldsymbol{p})}{c}, \nonumber \\
l_{a}(\boldsymbol{p}) &\equiv& \pi \frac{r(z_{\ast},\boldsymbol{p})}{r_{s}(z_{\ast},\boldsymbol{p})}\,,
\end{eqnarray}
in addition to $\Omega_b\,h^2$. Here $r_{s}(z_{\ast})$ is the comoving sound horizon evaluated at the photon-decoupling redshift evaluated using the fitting formula from \cite{Hu:1995en},
\begin{eqnarray}{\label{eq:zdecoupl}}
z_{\ast} &=& 1048 \left[ 1 + 0.00124 (\Omega_{b} h^{2})^{-0.738}\right] \times   \nonumber \\
&& \times \left(1+g_{1} (\Omega_{m} h^{2})^{g_{2}} \right)  \,,
\end{eqnarray}
where the factors $g_1$ and $g_2$ are given by
\begin{eqnarray}
g_{1} &=& \frac{0.0783 (\Omega_{b} h^{2})^{-0.238}}{1+39.5(\Omega_{b} h^{2})^{-0.763}}\,, \nonumber \\
g_{2} &=& \frac{0.560}{1+21.1(\Omega_{b} h^{2})^{1.81}} \,, \nonumber
\end{eqnarray}
while $r$ is the comoving distance at decoupling, i.e. from Eq.~(\ref{eq:comovdist}), $r(z_{\ast},\boldsymbol{p}) = D_M(z_{\ast},\boldsymbol{p})$.

{\renewcommand{\tabcolsep}{1.5mm}
{\renewcommand{\arraystretch}{2.}
\begin{table}
\begin{minipage}{0.48\textwidth}
\caption{Results from MCMC analysis. We report $1\sigma$ confidence intervals for each parameter and the Bayes Factors.}\label{tab:results1}
\centering
\resizebox*{\textwidth}{!}{
\begin{tabular}{ccccc}
\hline
   & \multicolumn{2}{c}{$\Lambda{\rm CDM}$} & \multicolumn{2}{c}{BH} \\
   & late & full & late & full \\
\hline
$\Omega_{ m}$   & $0.293^{+0.016}_{-0.016}$ & $0.319^{+0.005}_{-0.005}$    & $0.290^{+0.019}_{-0.019}$ & $0.317^{+0.007}_{-0.007}$  \\
$\Omega_{ b}$   & $-$                       & $0.0494^{+0.0004}_{-0.0004}$ & $-$                       & $0.045^{+0.002}_{-0.002}$ \\
$h$             & $0.713^{+0.013}_{-0.013}$ & $0.673^{+0.003}_{-0.003}$    & $0.715^{+0.014}_{-0.013}$ & $0.705^{+0.015}_{-0.015}$ \\
$\Delta$        & $-$                       & $-$                          & $>0.60$                   & $0.53^{+0.11}_{-0.07}$ \\
$C$             & $-$                       & $-$                          & $3.67^{+1.90}_{-1.79}$    & $2.13^{+0.63}_{-0.30}$ \\
\hline
$r_{s}(z_{\ast})$  & $-$                    & $143.81^{+0.19}_{-0.20}$     & $-$                       & $137.17^{+2.72}_{-2.61}$ \\
$r_{s}(z_{d})$     & $-$                    & $150.28^{+0.21}_{-0.22}$     & $-$                       & $143.25^{+2.86}_{-2.73}$ \\
$\mathcal{B}_{j}^{i}$           & $\mathit{1}$  & $\mathit{1}$   & $0.48^{+0.05}_{-0.05}$   & $32.604^{+0.001}_{-0.001}$ \\
$\ln \mathcal{B}_{j}^{i}$       & $\mathit{0}$  & $\mathit{0}$   & $-0.74^{+0.03}_{-0.03}$  & $3.48
^{+0.03}_{-0.04}$  \\
\hline
\end{tabular}}
\end{minipage}
\end{table}}}

{\renewcommand{\tabcolsep}{1.5mm}
{\renewcommand{\arraystretch}{2.}
\begin{table}
\begin{minipage}{0.48\textwidth}
\caption{Convergence test for MCMC. Convergence is achieved when all parameters have $j_{\ast}>20$ and $r<0.01$ \cite{Dunkley:2004sv}.}\label{tab:results2}
\centering
\resizebox*{\textwidth}{!}{
\begin{tabular}{ccccccccc}
\hline
   & \multicolumn{4}{c}{$\Lambda{\rm CDM}$} & \multicolumn{4}{c}{BH} \\
   & \multicolumn{2}{c}{late} & \multicolumn{2}{c}{full} & \multicolumn{2}{c}{late} & \multicolumn{2}{c}{full} \\
\hline
   & $j_{\ast}$ & $r$ & $j_{\ast}$ & $r$ & $j_{\ast}$ & $r$ & $j_{\ast}$ & $r$ \\
   & $(10^2)$ & $(10^{-3})$ & $(10^2)$ & $(10^{-3})$ & $(10^2)$ & $(10^{-3})$ & $(10^2)$ & $(10^{-3})$ \\
\hline
$\Omega_{ m}$   & $70$ & $0.2$ & $230$ & $0.2$ & $4$   & $0.6$ & $2$   & $2$ \\
$\Omega_{ b}$   & $-$  & $-$   & $10$  & $0.2$ & $-$   & $-$   & $0.8$  & $3$ \\
$h$             & $10$ & $0.2$ & $10$  & $0.2$ & $10$  & $0.4$ & $1$   & $3$ \\
$\Delta$        & $-$  & $-$   & $-$   & $-$   & $0.7$ & $2$   & $0.8$ & $3$ \\
$C$             & $-$  & $-$   & $-$   & $-$   & $12$  & $1$   & $0.4$ & $5$ \\
\hline
\end{tabular}}
\end{minipage}
\end{table}}}

\begin{figure*}[htbp]
\centering
\includegraphics[width=0.95\textwidth]{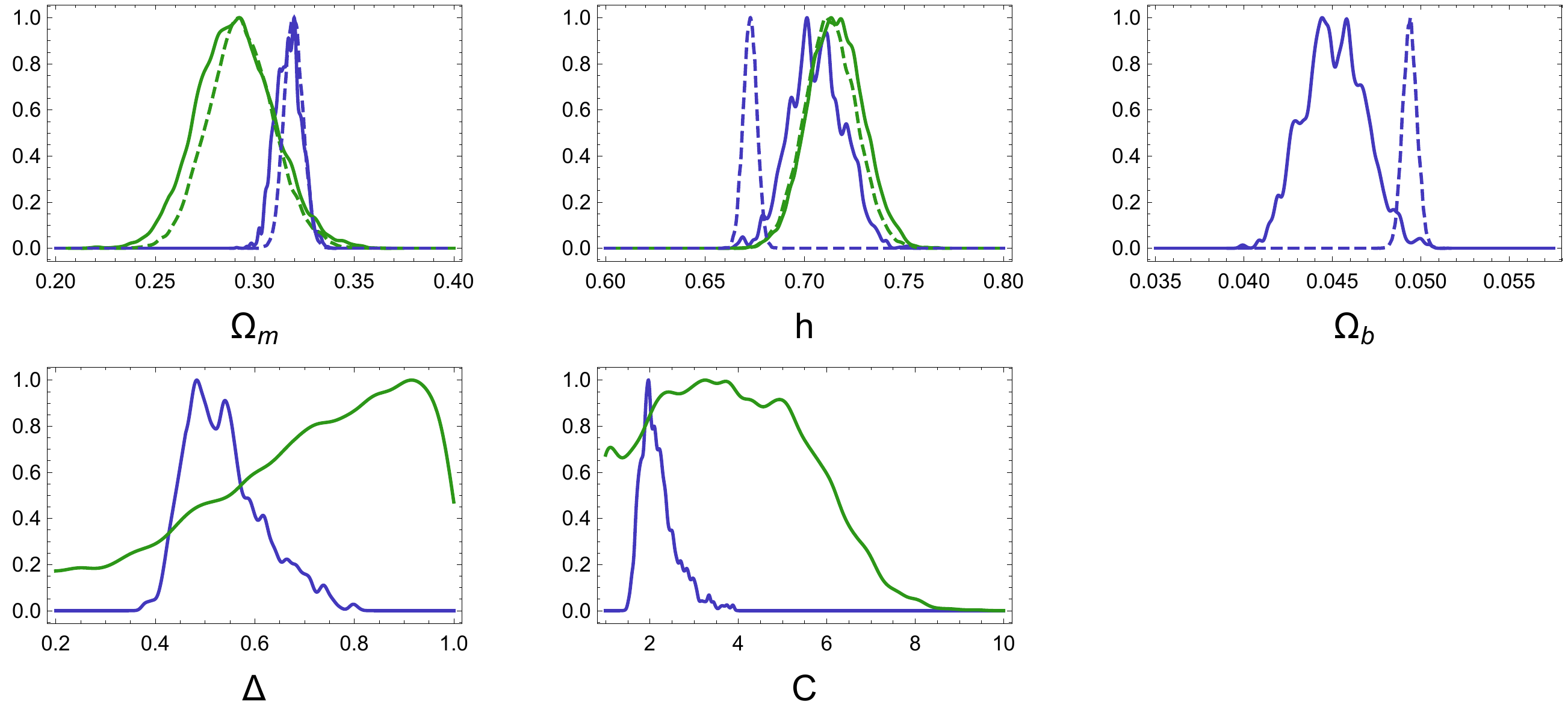}
\caption{Normalized histograms for each cosmological parameter. In green: late-time analysis; in blue: full cosmological data set. Dashed: $\Lambda$CDM model; solid: Barrow Holographic dark energy.}
\label{plot:1}
\end{figure*}

\begin{figure*}[htbp]
\centering
\includegraphics[width=0.95\textwidth]{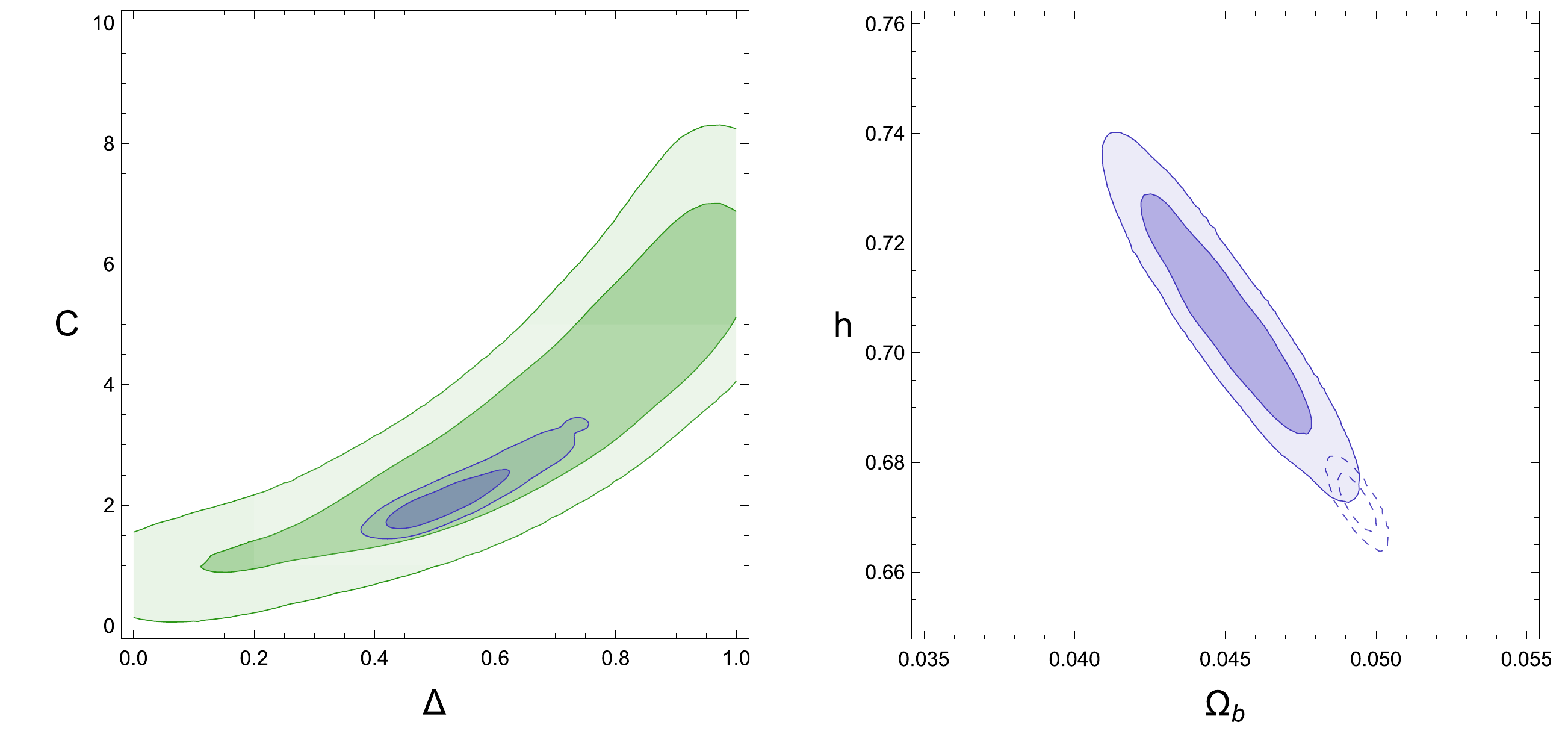}
\caption{Joint contours for the Barrow Holographic dark energy parameters \textit{(left panel)} and for the baryonic density parameter vs Hubble constant \textit{(right panel)}. In green: late-time analysis; in blue: full cosmological data set. Hard colors: $68\%$ confidence levels; soft colors: $95\%$ confidence levels. Dashed: $\Lambda$CDM model; solid: Barrow Holographic dark energy.}
\label{plot:2}
\end{figure*}

\begin{figure*}[htbp]
\centering
\includegraphics[width=0.95\textwidth]{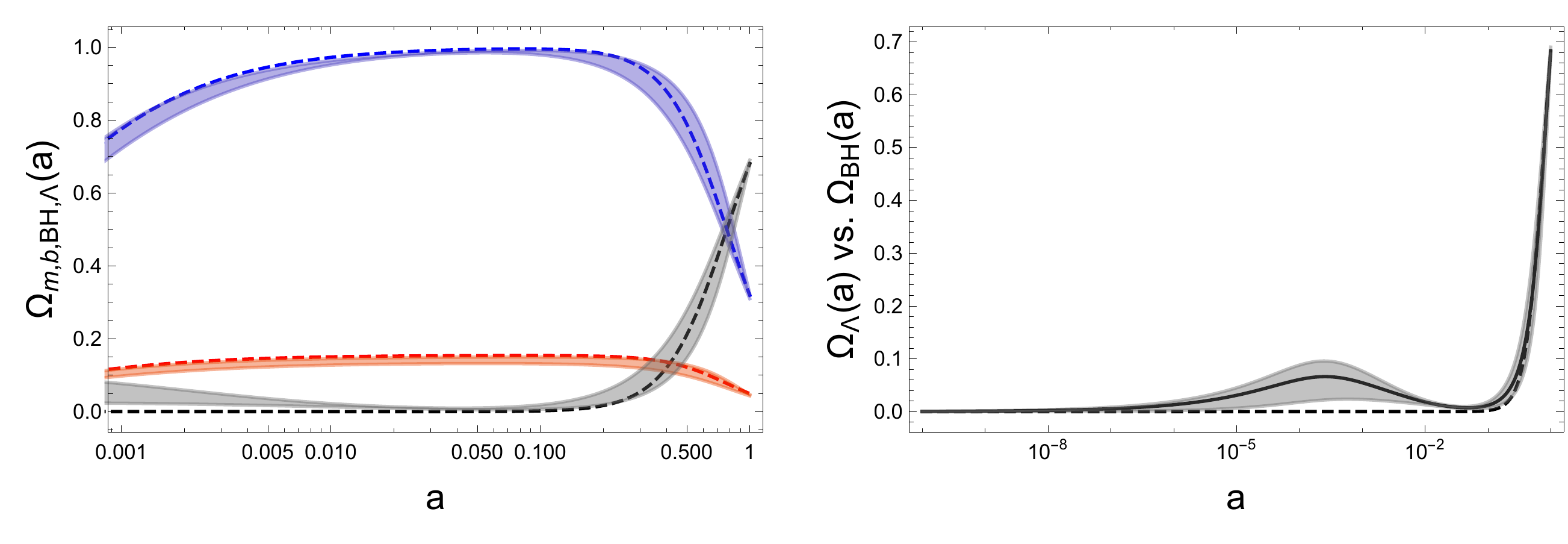}
\caption{Evolution with time of dimensionless density parameters. \textit{(left panel)} Matter (blue), baryons (red) and dark energy (grey/black) for both full BH model ($1\sigma$ confidence levels as shaded regions) and full $\Lambda$CDM (dashed lines). \textit{(right panel)}. Evolution of dark energy: full BH model ($1\sigma$ confidence levels as shaded regions and best fit as solid black line) vs full $\Lambda$CDM (dashed line).}
\label{plot:3}
\end{figure*}

\begin{figure*}[htbp]
\centering
\includegraphics[width=0.95\textwidth]{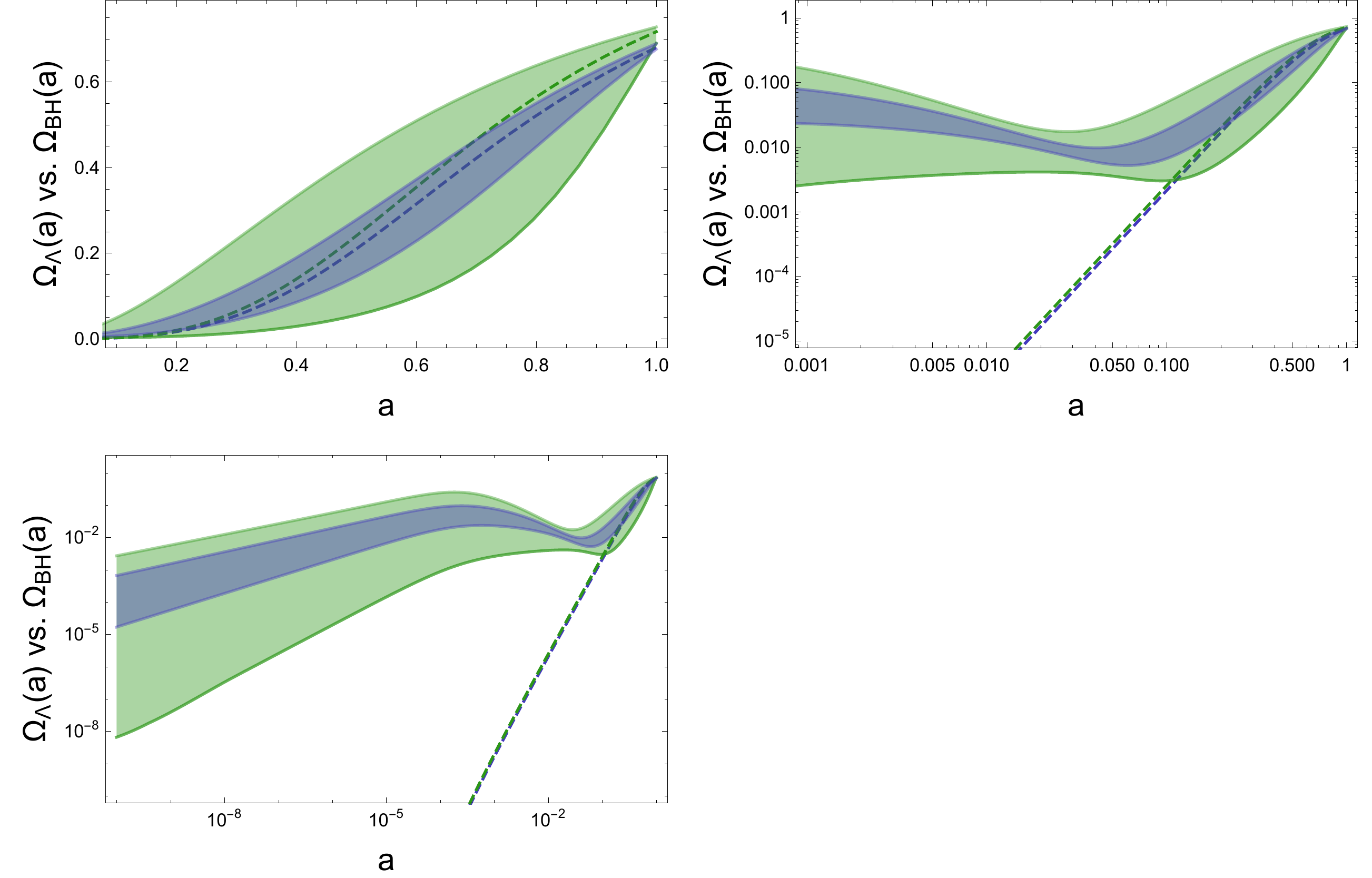}
\caption{Evolution of dark energy at different times. In green: late-time analysis; in blue: full cosmological data set. Dashed: $\Lambda$CDM model; shaded regions: Barrow Holographic dark energy ($1\sigma$ confidence levels).}
\label{plot:4}
\end{figure*}

\section{Discussion and Conclusions}

We display constraints on cosmological parameters from our MCMC analysis in Table~\ref{tab:results1}, and the convergence test on MCMC runs in Table~\ref{tab:results2}. The posterior distributions for each parameter are shown in Fig.~\ref{plot:1}.

If we compare our analysis to the one in \cite{Anagnostopoulos:2020ctz}, we can definitely assess that late-time-only observations are unable to constrain BH dark energy parameters, with the fractal parameter $\Delta$ spanning the full range of validity, and the BH characteristic energy scale $C$ being well defined (contrarily to what found in \cite{Anagnostopoulos:2020ctz}), but with large errors. The addition of early-time data {\em is crucial}: both parameters' confidence levels are now highly narrowed, as can be seen in the left panel of Fig.~\ref{plot:2}. Note that the value of the fractal parameter $\Delta$ totally rejects both the lower $(\Delta =0)$ and the upper $(\Delta = 1)$ limit.

The most striking result is given by the Bayes ratio: given Jeffreys' scale, there is ``strong evidence'', $\ln \mathcal{B}^{i}_{j} \approx 3.5$, in favour of BH dark energy w.r.t. a standard $\Lambda$CDM. This is a very surprising claim taking into account the outside of cosmological origin (i.e. COVID-19-like) nature of this type of dark energy, reinforced by the fact that the BH dark energy cannot be reduced to a cosmological constant.

Moreover, if we pay more attention to the values of the cosmological parameters, we can see that this statistical preference is led by a shift in both the Hubble constant, $h$, and the baryonic content, $\Omega_b$ (this is an important point, because $\Omega_b$ is called into question only when dealing with early-time data). As shown in the right panel of Fig.~\ref{plot:2}, we have a shift toward smaller values of $\Omega_b$, partially consistent with the $\Lambda$CDM scenario on the upper tail, and larger values of $h$, thus reducing the Hubble tension \cite{Freedman:2017yms,Riess:2019cxk,Riess:2020sih} to $\lesssim2.75\sigma$.

This shift in the cosmological parameters has as main consequence a shift in both the sound horizon value (at decoupling, $r_{s}(z_\ast)$ and at dragging, $r_{s}(z_d)$, epochs) and in the distance to last scattering. In Table~\ref{tab:results1} we report the values of both the sound horizons for all the considered models. The change with respect to $\Lambda$CDM is clear, but it is actually naturally expected if we look more carefully at the behaviour of the BH dark energy.

In Figs.~\ref{plot:3} and \ref{plot:4} we plot the evolution of the dimensionless density parameters, which clearly show how the BH dark energy model effectively behaves as an Early Dark Energy (EDE) model, although starting from totally different theoretical premises. The shifts in the geometrical distances connected to CMB physics are a quite known fact \cite{Linder:2008nq} for EDEs, and the Barrow holographic model perfectly fits this scenario.

In the left panel of Fig.~\ref{plot:3} we show the evolution with time of matter (blue), baryons (red) and dark energy (grey/black) for both full BH model ($1\sigma$ confidence levels as shaded regions) and full $\Lambda$CDM (dashed lines). For what concerns matter and baryons, the behaviours from the two scenarios are statistically equivalent both at late and early (approaching decoupling epoch) times, which are the regimes directly tested by our data. The departure at intermediate redshifts, $z \sim [10,100]$ is of course due to the lack of probes in that regime, but it still statistically not relevant.

Instead, for what concerns the dark energy component, we see that the fractal horizon-based holographic model from Barrow's proposal is an effective EDE, qualitatively and quantitatively. The right panel of Fig.~\ref{plot:3} compares BH model (the solid black line is the best fit-derived behaviour) with $\Lambda$CDM. The scales of the axis have been choosing for a better comparison with literature, as for example with Fig.~1 in \cite{Pettorino:2013ia}. In Fig.~\ref{plot:4}, we plot $\Omega_{BH}(a)$ in different redshift regimes, from left to right: late times only; till CMB (full redshift range of our probes); and at early times, prior to decoupling era.

In the right panel of Fig.~\ref{plot:3} we can note how we do have a bump in $\Omega_{BH}(a)$ at $z\sim 10^3\div10^4$ which, giving our results, is totally consistent with the used geometrical probes. Trying to have more insights about all these results, we can compare our work with literature for EDEs.

The \textit{Planck} survey has tried to put constraints on some EDE models \cite{Ade:2015rim}, with data from the $2015$ release (a more recent analysis with \textit{Planck} $2018$ data can be found in \cite{Colin:2020}). Although a direct comparison with the model considered here is not possible because the behaviours of the fluids do not resemble each other, we can have indirect insights in the physical soundness of BH dark energy if we calculate some parameters like, for example, the amount of dark energy at early time. For BH we have $\Omega_{BH}(a=0)\lesssim 4\cdot 10^{-4}$: even if we are using only geometrical probes, this constraint is as good as and as stringent as the one which can be derived by using the full power spectrum information from \textit{Planck}, see Table 3 and Fig.~11 in \cite{Ade:2015rim}.

A different EDE model has been analyzed in \cite{Poulin:2018cxd}; in their Table I one can find values for the sound horizon at recombination and check how for EDEs the shifts in such geometrical distances are natural, exactly like it happens for our model. More interestingly, the model in \cite{Poulin:2018cxd} is characterized by two further parameters: the scale factor $a_{c}$ at which we have a transition (or switch) in the EDE fluid from a cosmological constant-like behaviour to a dynamical one, and the amount of EDE expected at that same redshift. It turns out that the fit to the data (including \textit{Planck} $2015$ power spectrum) is consistent with $\log a_{c}\sim -3.7$ and an EDE fraction of $\sim5\%$. Our model, based only on geometrical probes, has the bump at $\log a_{c}\sim -3.6$ and the fraction of BH dark energy at that time is $\sim6.6\%$. We need to add also that the analysis in \cite{Poulin:2018cxd} uses at the same time both \textit{Planck} data and the measurement of $H_0$ from local probes, which is surely acting as a prior/bias on the final results, more specifically in lowering the Hubble tension. In our case, instead, we are not using any prior on $H_0$, but we have added data from H0LICOW. Thus, the agreement between the two scenarios, although very different, is quite strong, and we are confident that BH dark energy might be still successful in fitting data related to cosmological perturbations.

A further analysis about its influence on the growth of perturbations \cite{Mehrabi:2015kta} is of course needed and will be presented in a forthcoming paper. Careful attention will be needed to understand the role of dark energy perturbations, which are not taken into account in \cite{DAgostino:2019wko}, for example, but seem to be important to successfully fit the data, as shown in \cite{Poulin:2018cxd}. Thus the real question to be addressed will be \textit{only} how this further analysis will influence the new parameters of Barrow formulation (the fractal dimension in particular) and if the statistical preference for this model with respect to $\Lambda$CDM will stay untouched, lowered or improved.

\section*{acknowledgments}
We thank John Barrow for the constructive feedback and the anonymous Referee for helpful comments which helped to highlight new features of the model.

\end{document}